\documentclass[iop, apjs]{emulateapj}
\usepackage{threeparttable}
\usepackage{multirow}
\usepackage{times}
\usepackage{enumitem}
\usepackage{url} 
\usepackage{amsmath,bm}
\usepackage{color}

\shorttitle{}
\shortauthors{Y. Huang et al.}

\begin{document}

\title{A new catalogue of radial velocity standard stars from the APOGEE data}

\author{Y. Huang\altaffilmark{1,2,6}}
\author{ X.-W. Liu\altaffilmark{1}}
\author{ B.-Q. Chen\altaffilmark{1}}
\author{ H.-W. Zhang\altaffilmark{2,3}}
\author{ H.-B. Yuan\altaffilmark{4}}
\author{ M.-S. Xiang\altaffilmark{5,6}}
\author{ C. Wang\altaffilmark{2}}
\author{ Z.-J. Tian\altaffilmark{2,6}}
\altaffiltext{1}{South-Western Institute for Astronomy Research, Yunnan University, Kunming 650500, People's Republic of China; {\it yanghuang@pku.edu.cn {\rm (YH)}; x.liu@pku.edu.cn {\rm (XWL)}}}
\altaffiltext{2}{Department of Astronomy, Peking University, Beijing 100871, People's Republic of China}
\altaffiltext{3}{Kavli Institute for Astronomy and Astrophysics, Peking University, Beijing 100871, People's Republic of China}
\altaffiltext{4}{Department of Astronomy, Beijing Normal University, Beijing 100875, People's Republic of China}
\altaffiltext{5}{Key Laboratory of Optical Astronomy, National Astronomical Observatories, Chinese Academy of Sciences, Beijing 100012, People's Republic of China}
\altaffiltext{6}{LAMOST Fellow}

\begin{abstract}
We present a new catalogue of 18\,080 radial velocity standard stars selected from the APOGEE data. 
These RV standard stars are observed at least three times  and have a median stability ($3\sigma_{\rm RV}$) around $240$\,m\,s$^{-1}$ over a time baseline longer than 200 days.
They are largely distributed in the northern sky and could be extended to the southern sky by the future APOGEE-2 survey.
Most of the stars are red giants ($J - K_{\rm s} \ge 0.5$) owing to the APOGEE target selection  criteria. 
Only about ten per cent of them are main-sequence stars.
The $H$ band magnitude range of the stars is $7$--$12.5$\,mag with the faint limit much fainter than the magnitudes of previous RV standard stars.
As an application, we show the new set of standard stars to determine the radial velocity zero points  of the RAVE, the LAMOST {and the Gaia-RVS} Galactic spectroscopic surveys.

\end{abstract}

\section{Introduction}
The (barycentric or heliocentric) stellar radial velocity (RV) of a star is ideally defined as the change rate of distance between the Sun and the star.
It can be deduced from the {\it Doppler shift} of the spectrum of the star recorded in the reference frame of the telescope, and then transform the derived value to the barycentric or heliocentric reference frame. 
Accurate stellar RV measurements are of upmost importance for a variety of astrophysical studies, including Galactic kinematics and dynamics {(e.g. Bovy et al. 2012, 2015; Huang et al. 2015, 2016, 2017; Sun et al. 2015; Gaia Collaboration et al. 2018d), discoveries of stellar and substellar companions (e.g. binaries, brown dwarfs, and exoplanets; e.g. Mayor \& Queloz 1995; Udry \& Mayor 2008;  Gao et al. 2014, 2017; El-Badry et al. 2018; Tian et al. 2018) and stellar structure and evolution (pulsating variables, binaries and multiple systems; e.g. Chadid, Vernin \& Gillet 2008; Badenes \& Maoz 2012; Yang et al. 2012).
The typical accuracies for those various studies are required from few m\,s$^{-1}$ (e.g. for discovering of low-mass planets; Udry \& Mayor 2008) to few km\,s$^{-1}$ (for studying stellar structure and evolution, and Galactic kinematics and dynamics; Gao et al. 2014, 2017; Sun et al. 2015)}. 

In the past decades, the number of stellar RV measurements has increased dramatically, thanks to several either already completed or still ongoing large-scale spectroscopic surveys, e.g. the RAVE (Steinmetz et al. 2006), the SDSS/SEGUE (Yanny et al. 2009), the SDSS/APOGEE (Majewski et al. 2017), the LAMOST (Deng et al. 2012; Liu et al. 2014), the Gaia-ESO (Gilmore et al. 2012), the HERMES-GALAH (De Silva et al. 2015), and the Gaia-RVS (Katz 2009; {Katz et al. 2018}) surveys.
More ambitious large-scale spectroscopic surveys are under-plan and upcoming, such as the WEAVE (Dalton et al. 2014) and the 4MOST ( de Jong et al. 2016) surveys.

To achieve accurate RV measurements, it is essential to build sets of RV standard stars suitable for various spectroscopic surveys.
High quality RV standard stars are needed to determine the RV zero-points (RVZPs) of the instruments employed by various surveys.
As pointed by Crifo et al. (2010), the concept of RV standard stars is based on the physical notion ``stability'' rather than on other more fundamental physical definitions (e.g. Lindegren \& Dravins 2003).
As such, the definition of RV standard stars can change depending on the accuracy of RV measurements expected for the surveys.

Up till now, there are a few thousand RV standard stars with velocity uncertainties less than 100\,m\,s$^{-1}$ over a time baseline longer than one year, owning to efforts of several groups (e.g. Udry, Mayor \& Queloz 1999, hereafter UMQ99; Nidever et al. 2002, hereafter N02; Chubak et al. 2012, hereafter C12; Soubiran et al. 2013, hereafter S13; {Soubiran et al. 2018}).
The stability of those stars are good enough for almost all existing/planning spectroscopic surveys.
However, the limited number as well as the relatively bright limiting magnitude ($V \sim$\,9--10\,mag, see Section\,2) of the  stars prevent reliable determinations of RVZPs for most of the existing/planning surveys given that only a very limited number of standard stars are targeted by the surveys.

To solve the problem, we attempt to construct a new set of RV standard stars from the APOGEE data based on the following considerations.
First, the high resolution ($R$\,$\sim$\,22\,500) and high signal-to-noise ratios (SNRs) near-infrared ($H$ band; 1.51\,--\,1.70\,$\mu$m)  spectra collected by SDSS/APOGEE deliver stellar RV measurements of a precision of $\sim$\,$100$\,m\,s$^{-1}$ (Holtzman et al. 2015).
Secondly,  the primary science targets selected by the survey lie in the magnitude range between $7 \le H \le 13.8$\,mag.
The bright limit of this range yields sufficient number of stars in common with the existing RV standard stars allowing a robust determination of the RVZP of the APOGEE survey itself, whereas the faint limit ensures that a large number of  APOGEE RV standard stars thus built can be targeted by other spectroscopic surveys.
Thirdly, the recently released SDSS\,DR14 (Abolfathi et al. 2017) includes APOGEE observations of  over one million spectra of 277\,371 unique stars collected by APOGEE-1 (September\,2011-July\,2014) and APOGEE-2 (July\,2014-July\,2016) surveys. 
Over 80 per cent  of the 277\,371 unique stars have been visited by two times or more. 
The very high fraction ($\ge 80$\%) of multi-epoch observations for a large number (over two hundred thousand) of stars with a long time baseline (nearly five years) allows one to select and build a large sample of RV standard stars.

The paper is organized as follows. 
In Section\,2, we describe the reference RV standard stars selected from the existing databases.
In Section\,3 we describe in detail the construction of a sample of RV standard stars from the APOGEE data.
As an application, we use the new set of  standard stars to examine the RVZPs of several finished/ongoing large-scale spectroscopic surveys in Section\,4.
We summarize in Section\,5.

\section{Reference RV standard stars from the existing databases}
In this Section, we attempt to build a set of reference RV standard stars selected from the existing databases.
These reference RV standard stars will be used to calibrate the RV measurements of the APOGEE survey in the next Section.
In doing so, we first briefly introduce the four databases used in the current work as follows.

The first one is the 38 ``New'' ELODIE-CORAVEL high-precision standard stars (UMQ99), the official set of IAU RV standards.
These stars have been observed more than $\sim$\,10 times for years and have velocity variations lower than a few 10 m\,s$^{-1}$.

The second one, provided by N02, contains 889 late-type stars observed with the HIRES echelle spectrometer on the 10\,m Keck\,I telescope  and with the ``Hamilton'' echelle spectrometer installed on either the 3\,m Shane telescope or the 0.6\,m Coude Auxilliary Telescope (CAT).
All those stars are typically visited by 12 times in the period between 1997 and 2001 and 782 of them show velocity variations smaller than 100\,m\,s$^{-1}$.
 
The third one, built by C12, is an extension of the N02 database to include more stars (2046 FGKM-type stars in total) observed with the HIRES echelle spectrometer on the 10\,m Keck\,1 telescope between August 2004 and January 2011, as parts of the California Planet Survey (CPS; Howard et al. 2010). 
Amongst the 2046 FGKM-type stars, 131 are selected as RV standard stars by C12 since they exhibit stable RV for at least 10 years, with a velocity scatter less than 30\,m\,s$^{-1}$.

Finally, S13 provide a catalogue of 1420 potential RV standard stars with data taken from either archives or new observations. 
Over $90$\% of them exhibit a stability better than 300\,m\,s$^{-1}$ over several years.  
The main goal of the project is to provide a stable RV reference stars to calibrate RV measurements of Gaia-RVS.
The project uses data collected with five high-resolution spectrographs: ELODIE ($R = 42\,000$), SOPHIE ($R = 75\,000$), NARVAL ($R = 78\,000$), CORALE ($R = 50\,000$) and HARPS ($R = 120\,000$).
RV measurements from the different spectrographs are combined, adopting the SOPHIE measurements as the reference frame. 
The results are compared to values published in other existing catalogues, including N02 and C12.
The RV offset between S13 and UMQ99 values reflects  the RVZP  between SOPHIE and ELODIE frames, and is also reported in S13. 

Following S13, we adopt the  SOPHIE frame as the RV reference scale when combining the four aforementioned RV databases.
The corrections used to correct measurements of the other three databases to S13 (SOPHIE frame) are again taken from S13 and listed in Table\,1.
The criteria for the selection of the final reference RV standard stars are $N_{\rm obs} \ge 3$;  time baseline longer than one year and stability ($3\sigma_{\rm RV}$) better than $100$\,m\,s$^{-1}$.
In case a star is available  from more than two databases, the preference is that S13 first, UQM99 second and C12 third.
Finally, we obtain a total of  1611 reference RV standard stars, with 29 from UMQ99, 347 from N02, 332 from C12 and 903 from S13 (see Table\,1). 
The spatial distribution of these stars is shown in Fig.\,1, and is quite homogeneous across the whole sky.
The color-magnitude ($V$ against $B - V$) of the stars is presented in Fig.\,2.
As the plot shows, most of them are FGK-type stars brighter than 10\,mag and of colors between 0.5 and 1.0\,mag.
The sample also includes a few hundred M-type stars fainter than 10\,mag and of colors redder than 1.0\,mag.

\begin{figure}
\begin{center}
\includegraphics[scale=0.4,angle=0]{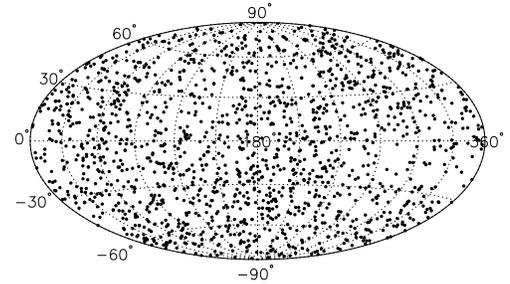}
\caption{Distribution of the 1611 reference RV standard stars on the celestial sphere in the Galactic coordinate system.}
\end{center}
\end{figure}

\begin{figure}
\begin{center}
\includegraphics[scale=0.4,angle=0]{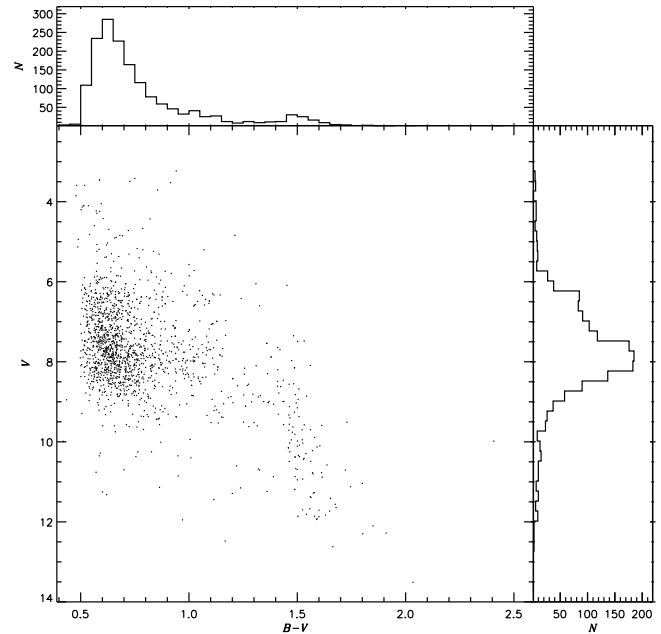}
\caption{$V$ versus $B-V$ color-magnitude diagram of the 1611 reference RV standard stars.
Histogram distributions along the axes $V$ and $B - V$ are also plotted.}
\end{center}
\end{figure}

\begin{table}
\centering
\caption{Selection of reference  radial velocity standard stars from the existing databases}
\begin{threeparttable}
\begin{tabular}{lccc}
\hline
Database & $N_{\rm obs}$ & $N_{\rm sele}$ & RVZP correction (m\,s$^{-1}$)\\
\hline
UMQ99&38&29& $−259.0 (B - V) + 105.2$\\

\multirow{2}{*}{N02}&\multirow{2}{*}{889}&\multirow{2}{*}{347}&$72$\tnote{a}\\
&&&$-141$\tnote{b}\\

\multirow{2}{*}{C12}&\multirow{2}{*}{2177}&\multirow{2}{*}{332}&$63$\tnote{a}\\
&&&$-98$\tnote{b}\\

S13&1420&903&$0$\\
\hline
\end{tabular}
\begin{tablenotes}
\item[a] For RV measurements using the template of the Sun.
\item[b] For RV measurements using the template of an M dwarf. 
\end{tablenotes}
\end{threeparttable}
\end{table}

\section{APOGEE Radial velocity standard stars}
\subsection{The APOGEE survey}
The APOGEE survey currently includes APOGEE-1 and APOGEE-2.
As an important component of SDSS-III (Eisenstein et al. 2011), the APOGEE-1 survey was executed from September 2011 to July 2014 using the APOGEE-North spectrograph on the Sloan Foundation 2.5\,m telescope of Apache Point Observatory (APO). 
It has collected  about half million high SNR, high resolution ($R \sim 22\,500$), near-infrared ($H$ band; 1.51\,--\,1.70\,$\mu$m) spectra  for over 163\,000 stars in the bulge, disk and halo of our Galaxy.
The target selections and scientific motivations of APOGEE-1 are described in Zasowski et al. (2013) and Majewski et al. (2017), respectively.
As a successor of APOGEE-1 and part of SDSS-IV (2014--2020; Blanton et al. 2017), the APOGEE-2 survey will expand the APOGEE-1 to an all-sky $H$-band spectroscopic survey using the original APOGEE-North spectrograph on the Sloan Foundation 2.5\,m telescope of APO and another clonal spectrograph (APOGEE-South) on the Ir{\'e}n{\'e}e du Pont 2.5m telescope of Las Campanas Observatory (LCO).
The survey expects to collect high resolution, near-infrared spectra of $\sim 3 \times 10^{5}$ stars across the entire sky (Majewski et al. 2016; Zasowski et al. 2017).
Benefited from the high resolution and high SNR spectra, the APOGEE survey provides RV measurements of a precision around $100\,$m\,s$^{-1}$ and a zero-point uncertainty at the level of $500\,$m\,s$^{-1}$ (Nidever et al. 2015), estimates of fundamental parameters with an accuracy better than 150\,K for $T_{\rm eff}$, 0.2\,dex for log\,$g$  and 0.1\,dex for [Fe/H] (M{\'e}sz{\'a}ros et al. 2013; Garc{\'{\i}}a P{\'e}rez et al. 2016), and estimates of up to 15 chemical elements with a typical precision of 0.1\,dex ( Garc{\'{\i}}a P{\'e}rez et al. 2016).

In the current work, we use the APOGEE data released in SDSS\,DR14 (Abolfathi et al. 2017), which contains about one million spectra of around thirty thousand unique stars collected during APOGEE-1 (2011-2014) and APOGEE-2 (2014-2016).

\subsection{RVZP of the APOGEE instrument}
Before selecting RV standard stars from the APOGEE data, we first examine the RVZP of the APOGEE instrument using the reference RV standard stars constructed in Section\,2.
In doing so, we cross match the reference RV standard stars with the APOGEE DR14\footnote{https://data.sdss.org/sas/dr14/apogee/spectro/redux/r8/stars/l31c/l31c.2/\\allStar-l31c.2.fits} catalog of stellar properties (including RV) deduced from the combined spectra and find 76 stars in common.

Generally, the RVZP of an instrument is affected by a number of factors, including the instrument itself, the spectral resolution and coverage, the wavelength calibration and the method of deriving RV (e.g. by cross-correlation), and observing conditions and environment {(e.g. Gullberg \& Lindegren 2002; S13)}.
Provided the environmental conditions are well controlled, the offset between two instruments is typically a constant value or shows some dependencies on the stellar color (or effective temperature), caused by the variations of spectral features used to derive RVs.
We therefore examine the offset between the APOGEE and the reference RV measurements as a function of effective temperature ($T_{\rm eff}$), as measured from the APOGEE spectra (Fig.\,3).
As the plot shows, the offset shows a clear, nearly linear trend of variations with effective temperature.
The offset is about $0.5$\,km\,s$^{-1}$ at $T_{\rm eff} \sim 6000$\,K and $-0.2$\,km\,s$^{-1}$ at $T_{\rm eff} \sim 3500$\,k.
To quantitively describe the trend, a simple linear fit is applied as follows:

\begin{equation}
 {\Delta {\rm RV}} = - 1.1069 + 0.2558 \times (T_{\rm eff}/10^{3} {\rm\,K}) {\rm\,km\,s^{-1}},
\end{equation}
where $\Delta {\rm RV}$ is ${\rm RV_{APOGEE}} - {\rm RV_{REF}}$.
This equation will be used to correct the RVs of APOGEE RV standard stars in the next Section.

{The origin of the effective temperature (color) dependence of the offset between the APOGEE RVs and the RVs from the reference standard stars is quite complicate.
One  probable cause is that their RVs are derived from spectra with different wavelength coverage, namely, the RVs from the APOGEE are derived using near-infrared spectra while the RVs from the reference standard stars are derived using optical spectra.}

\begin{figure}
\begin{center}
\includegraphics[scale=0.4,angle=0]{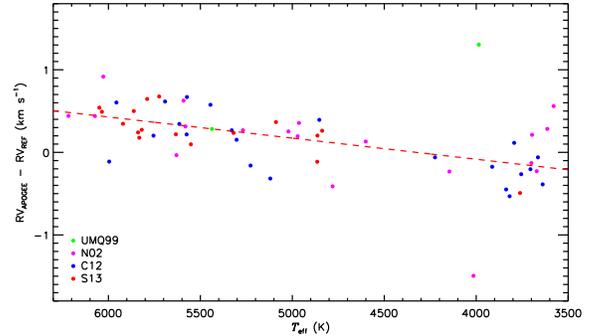}
\caption{Differences of APOGEE (RV$_{\rm APOGEE}$)  RV measurements and reference values of RV standard stars (RV$_{\rm REF}$)  as a function of effective temperature ($T_{\rm eff}$).
Dots of different colors represent measurements from different databases (as labelled in the left bottom corner of the plot) of the reference RV standard stars.}
\end{center}
\end{figure}

\subsection{Selecting RV standard stars from the APOGEE data}
\begin{figure*}
\begin{center}
\includegraphics[scale=0.425,angle=0]{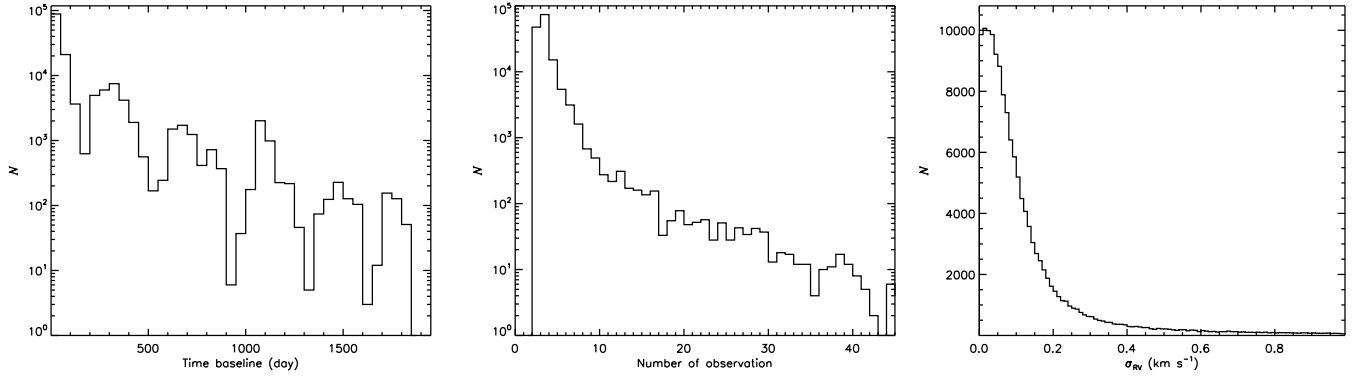}
\caption{Distributions of time baseline (left panel), number of observations (middle panel) and $\overline{\rm RV}$ wighted standard deviation ($\sigma_{\rm RV}$, right panel) of stars with at least two high-quality (SNRs\,$\ge 50$ and $\sigma_{\rm RV} \le 500$\,m\,s$^{-1}$) observations from the APOGEE data.}
\end{center}
\end{figure*}

\begin{figure*}
\begin{center}
\includegraphics[scale=0.425,angle=0]{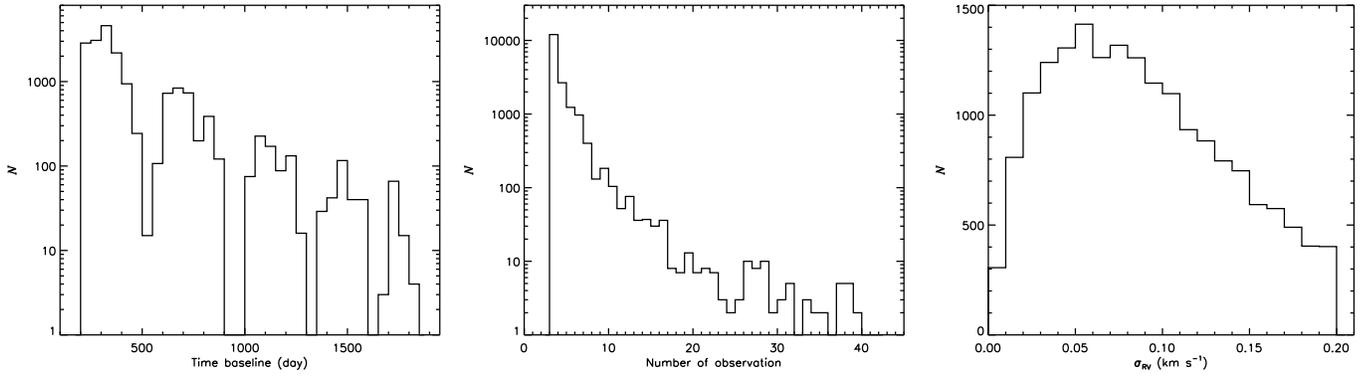}
\caption{Similar to Fig.\,5 but for the final sample of APOGEE RV standard stars (see Section\,3.3 for definitions).}
\end{center}
\end{figure*}

In this section, we attempt to select RV standard stars from the APOGEE data.
To do so, we first combine APOGEE RV measurements from the individual spectra obtained in multiple visits as published in SDSS DR14\footnote{https://data.sdss.org/sas/dr14/apogee/spectro/redux/r8/stars/l31c/l31c.2/\\allVisit-l31c.2.fits}.
Only results from spectra of SNR greater than 50 with RV measurement error smaller than 500\,m\,s$^{-1}$  are included in the combination.
Following S13, we define the following quantities when combining the data:
\begin{itemize}[leftmargin=*]
\item Weighted mean RV: $\overline{\rm RV} = \frac{{\sum\limits_{i=1}^{n}\rm RV_{i}}w_{i}}{\sum\limits_{i = 1}^{n}w_{i}}$, where $w_{i}$ is weight as given by the individual RV measurement error: $1/\epsilon_{i}^{2}$ and $n$ is the total number of observations;
\item Internal error of  $\overline{\rm RV}$: $I_{\rm ERV} = \frac{{\sum\limits_{i=1}^{n}\rm \epsilon_{i}}w_{i}}{\sum\limits_{i = 1}^{n}w_{i}}$;
\item $\overline{\rm RV}$ wighted standard deviation:

 $\sigma_{\rm RV}^{2} = \frac{\sum\limits_{i=1}^{n}w_{i}}{(\sum\limits_{i=1}^{n}w_{i})^{2}-\sum\limits_{i=1}^{n}w_{i}^{2}}\sum\limits_{i}^{n}w_{i}({\rm RV_{i}} - \overline{\rm RV})^{2}$;
 
 \item Uncertainty of $\overline{\rm RV}$: max($\sigma_{\rm RV}/\sqrt{N}$, $I_{\rm RV}/\sqrt{N}$);
 
 \item Time baseline in days: $\Delta T$;
 
 \item Mean Modified Julian day (MJD) of the $n$ observations.
\end{itemize}
A total of about 150 thousand stars are selected with at least two high-quality (i.e. SNRs\,$\ge 50$ and $\sigma_{\rm RV} \le 500$\,m\,s$^{-1}$) observations from the APOGEE data.
The distributions of time baseline ($\Delta T$), number of observations ($n$) and $\overline{\rm RV}$ wighted standard deviation ($\sigma_{\rm RV}$) of the selected stars are presented in Fig.\,5.
As the plot shows, most stars have $\Delta T$ smaller than 200 days (close to the period of one observational season) and $n$ smaller than 5.

\begin{figure}
\begin{center}
\includegraphics[scale=0.4,angle=0]{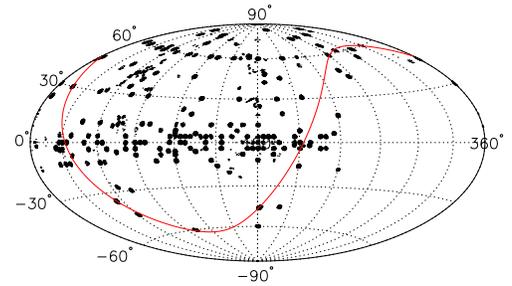}
\caption{Spatial distribution of the APOGEE RV standard stars on the celestial sphere in the Galactic coordinate system.
              The red line delineates the celestial equator.}
\end{center}
\end{figure}

\begin{figure}
\begin{center}
\includegraphics[scale=0.4,angle=0]{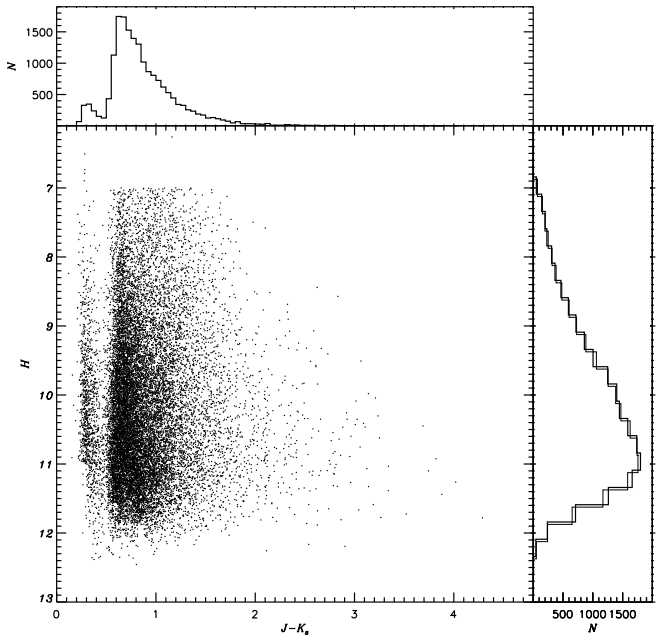}
\caption{$H$ versus $J - K_{\rm s}$ color magnitude diagram of the final sample of APOGEE RV standard stars.
Histogram distributions along the axes of $J$ and $J - K_{\rm s}$ are also plotted.}
\end{center}
\end{figure}

\begin{table*}
\centering
\caption{The final sample of APOGEE radial velocity standard stars}
\begin{threeparttable}
\begin{tabular}{ccccccccccc}
\hline
Name & $H$ & $J-K_{\rm s}$ & $T_{\rm eff}$ & $\overline{\rm RV}$\tnote{a} & $I_{\rm ERV}$ & $\sigma_{\rm RV}$ & $n$ & Uncertainty & $\Delta T$ & Mean MJD \\
&&&(K)&(km\,s$^{-1}$)&(km\,s$^{-1}$)&(km\,s$^{-1}$)&&(km\,s$^{-1})$&(days)&($-50000$)\\
\hline
  J00:00:00.68+57:10:23.4&10.13& 0.65&4987&$ -12.335$&0.0092&0.1336& 3&0.0772& 776&6389\\
  J00:00:21.18+61:36:42.1&10.18& 1.27&4214&$-122.852$&0.0059&0.1142& 5&0.0511& 274&7211\\
  J00:00:24.72+55:18:47.3&10.70& 0.98&4298&$-128.941$&0.0118&0.0940& 3&0.0543& 776&6389\\
  J00:00:25.61+55:28:51.1&11.20& 0.50&5512&$ -41.319$&0.0209&0.1276& 3&0.0737& 776&6389\\
  J00:00:31.19+70:56:36.5&10.17& 1.09&4300&$ -55.158$&0.0047&0.0612& 3&0.0353& 322&5964\\
  J00:00:34.75+57:23:25.9&11.57& 0.47&5657&$ -17.006$&0.0212&0.1584& 3&0.0914& 776&6389\\
  J00:00:46.32+56:34:05.7& 9.83& 0.75&4727&$ -11.147$&0.0080&0.0658& 3&0.0380& 776&6389\\
  J00:00:48.73+56:47:03.1&11.74& 0.80&4759&$ -73.246$&0.0158&0.0568& 3&0.0328& 776&6389\\
  J00:00:51.43+56:15:56.9&10.94& 0.69&4912&$ -60.790$&0.0114&0.0273& 3&0.0158& 776&6389\\
  J00:00:55.91+63:05:05.1& 9.92& 1.08&4367&$-108.744$&0.0072&0.1898& 5&0.0849& 274&7211\\
                                                       ...&...&...&...&...&...&...&...&...&...&...\\
  J05:50:45.34+17:53:46.6& 9.55& 0.72&4787&$  35.126$&0.0052&0.1103& 4&0.0552& 281&6432\\
  J05:50:46.37+11:07:47.7& 9.48& 0.81&3601&$ -11.462$&0.0042&0.1199& 9&0.0400& 393&6510\\
  J05:50:47.95-03:54:34.5& 9.95& 0.78&4932&$  90.759$&0.0061&0.1586& 3&0.0916& 743&6135\\
  J05:50:49.38-04:26:16.8&11.07& 0.77&4969&$  61.452$&0.0151&0.0725& 3&0.0418& 743&6135\\
  J05:50:50.72+16:53:37.3&10.46& 0.73&4913&$  74.542$&0.0133&0.0574& 3&0.0331& 281&6388\\
  J05:50:54.43+26:10:24.0&10.77& 1.01&5116&$  37.375$&0.0085&0.0746& 5&0.0333& 379&6885\\
  J05:50:55.24+17:52:55.3& 9.08& 1.05&4182&$  76.459$&0.0029&0.1132& 4&0.0566& 281&6432\\
  J05:50:55.60+17:47:35.5&10.31& 0.80&4664&$  30.372$&0.0077&0.0995& 4&0.0497& 281&6432\\
  J05:51:02.96+52:39:15.3&10.73& 0.72&4893&$ -46.840$&0.0111&0.0891& 3&0.0514& 237&7146\\
  J05:51:04.50+52:24:28.5& 9.43& 0.72&4887&$ -19.496$&0.0049&0.0290& 3&0.0168& 237&7146\\
                                                       ...&...&...&...&...&...&...&...&...&...&...\\
\hline
\end{tabular}
\begin{tablenotes}
\item[a] After applying the RVZP corrections\ given by Eq.\,(1).

(The full table is available in a machine-readable form in the online journal. A portion of it is shown here for illustration purpose only.)
\end{tablenotes}
\end{threeparttable}
\end{table*}

\begin{table*}
\centering
\caption{Comparisons of RVs yielded by the RAVE and LAMOST pipelines and those of the APOGEE RV standard stars}
\begin{threeparttable}
\begin{tabular}{lcccc}
\hline
Source & $\Delta {\rm RV}$ (km\,s$^{-1}$) & s.d. (km\,s$^{-1}$) & $\overline{\rm SNR}$& $N$ \\
\hline
RAVE (Siebert et al. 2011)&$+0.17$&$1.27$&$48$&$352$\\
LAMOST (LSP3; Xiang et al. 2015, 2017)&$-2.60$&$3.86$&$60$&$4849$\\
LAMOST (LASP; Luo et al. 2015)&$-3.92$&$3.63$&$46$&$3290$\\
{Gaia-RVS (Katz et al. 2018)}& {$+0.36$}& {$0.59$}&{--}&{$10674$}\\
\hline
\end{tabular}
\end{threeparttable}
\end{table*}

We further refine the sample of APOGEE RV standard stars with following criteria:
\begin{itemize}[leftmargin=*]
\item At least 3 measurements available from high-quality (SNRs\,$\ge 50$ and $\sigma_{\rm RV} \le 500$\,m\,s$^{-1}$) observations;
\item $\overline{\rm RV}$ wighted standard deviation $\sigma_{\rm RV} \le 200$\,m\,s$^{-1}$ (i.e. stability better than 600\,m\,s$^{-1}$);
\item Time baseline $\Delta T$ longer than 200 days.
\end{itemize}
{Here, we note that the RV stability is defined as $3\sigma_{\rm RV}$.}
A total of 18\,080 APOGEE RV standard stars are selected fulfilling the above criteria.
The distributions of time baseline ($\Delta T$), number of observations ($n$) and $\overline{\rm RV}$ wighted standard deviation ($\sigma_{\rm RV}$) of these RV standard stars are presented in Fig.\,6.
They have  a median $\sigma_{\rm RV}$ of 80\,m\,s$^{-1}$ (i.e. a median stability of 240\,m\,s$^{-1}$), good enough for the calibration of the RVZPs of most of the currently finished/ongoing  large-scale Galactic spectroscopic surveys (e.g. SDSS/SEGUE, RAVE, LAMOST).
Their spatial distribution is shown in Fig.\,6.
Most of them stars are in the northern sky given that most of the observations are collected at APO.
When the observations of APOGEE-2 collected at LCO become more and more, we expect more stars in the south.
In Fig.\,7 we show the color-magnitude ($H$ against $J - K_{\rm s}$) of this final sample of RV standard stars.
Most stars ($\ge 90$\%) are redder than $0.5$ in $J - K_{\rm s}$ due to the APOGEE target selection algorithm (Zasowski et al. 2013, 2017).
About ten per\,cent bluer ($J - K_{\rm s} \le 0.5$\,mag) main-sequence stars are also available, as add-on targets in the APOGEE survey.
The cut at faint end ($\sim 12$--$12.5$\,mag) in $H$ band magnitudes of these standard stars is brighter than the limiting magnitude ($H \sim 13.8$\,mag) of the APOGEE survey simply because of our SNR cut ($\ge 50$) when selecting the stars.

Finally, we correct the values of  $\overline{\rm RV}$ of these stars for RVZP using Eq.\,(1) and $T_{\rm eff}$ given by the combined spectra.
Properties of the sample stars, including name, $H$ magnitude, $J-K_{\rm s}$, $T_{\rm eff}$,  $\overline{\rm RV}$, $I_{\rm ERV}$, $\sigma_{\rm RV}$, $n$,  uncertainty of $\overline{\rm RV}$, $\Delta T$ and mean MJD, of the final sample of 18\,080 APOGEE RV standard stars are listed in Table\,2.

\begin{figure*}
\begin{center}
\includegraphics[scale=0.45,angle=0]{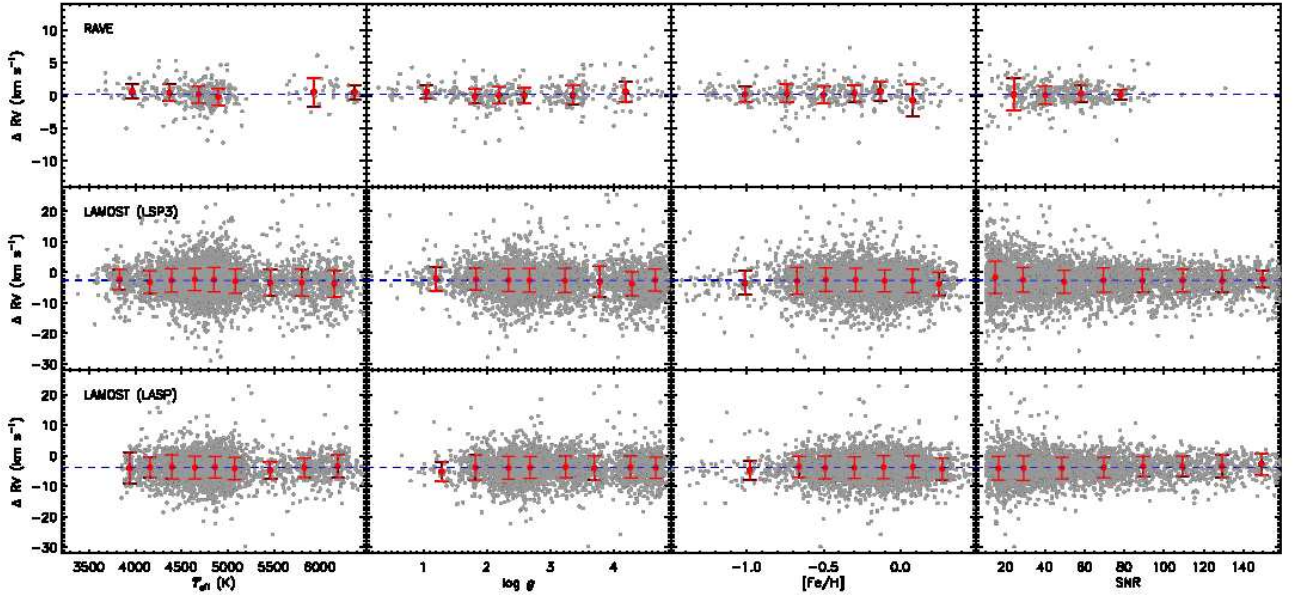}
\caption{Differences of RVs yielded by the RAVE (top panel), LAMOST/LSP3 (middle panel), and LAMOST/LASP (bottom panel) pipelines and those of the APOGEE RV standard stars are plotted as a function of $T_{\rm eff}$ (first column), log\,$g$ (second column), [Fe/H] (third column) and SNR (last column).
The red dots and error bars in the sub-panel represent the means and standard deviations of the differences in the individual parameter ($T_{\rm eff}$, log\,$g$, [Fe/H] and SNR) bins.
Blue dash lines delineate the mean differences presented in Table\,3.}
\end{center}
\end{figure*}

\begin{figure*}
\begin{center}
\includegraphics[scale=0.45,angle=0]{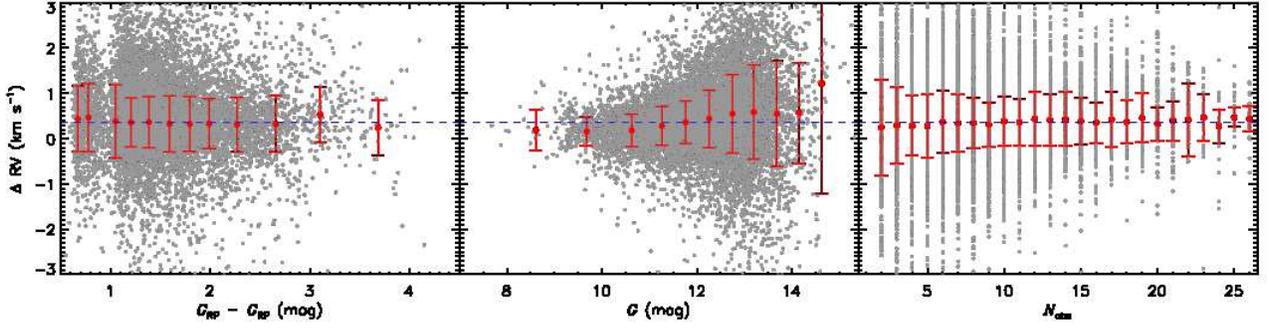}
\caption{Differences of RVs yielded by the Gaia-RVS pipeline and those of the APOGEE RV standard stars are plotted as a function of color $G_{\rm BP} -G_{\rm RP}$ (left panel), magnitude $G$ (middle panel) and  number of transits $N_{\rm obs}$ (right panel).
The red dots and error bars in each panel represent the means and standard deviations of the differences in the individual parameter ( $G_{\rm BP} -G_{\rm RP}$, $G$  and $N_{\rm obs}$) bins.
Blue dash lines delineate the mean differences presented in Table\,3.}
\end{center}
\end{figure*}

\begin{figure*}
\begin{center}
\includegraphics[scale=0.45,angle=0]{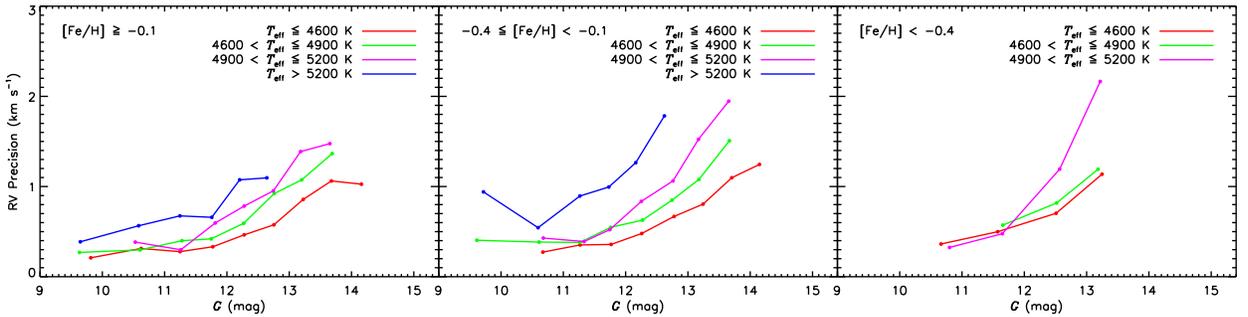}
\caption{RV precision of Gaia-RVS as function of $G$ magnitude reported by comparisons to the APOGEE RV standard stars, respectively for metal-rich stars ([Fe/H]\,$\ge -0.1$, left panel), middle metallicity stars ($-0.4 \leq$\,[Fe/H]\,$<-0.1$, middle panel) and relatively metal-poor stars ([Fe/H]\,$ < -0.4$, right panel).
The curves have been calculated for different APOGEE effective temperature ranges as labelled in the right-top cornel of each panel.}
\end{center}
\end{figure*}

\section{Calibrations of radial velocity scales for stellar spectroscopic surveys}
As mentioned, RVZPs of large-scale stellar spectroscopic surveys need to be determined and corrected for further studies (e.g. Galactic kinematics and dynamics).  
In this Section, we attempt to calibrate the radial velocity measurements by determining their RVZPs for {three recent stellar spectroscopic surveys: the RAVE, the LAMOST Galactic Spectroscopic Surveys and the Gaia-RVS survey}, using the new APOGEE RV standard star catalogue constructed above.
The results are presented in Figs.\,8, and 9, and Table\,3, including the median offset $\Delta RV$ of RVs measured by the two surveys and those given by the APOGEE RV standard star catalogue, the standard deviation of the offset, s.d., the median spectral SNRs of stars used in the comparison, and the number of stars $N$ used. 

\textbf{(i) The RAVE survey:} The survey (Steinmetz et al. 2006; Kunder et al. 2017) has collected 520\,781 medium-resolution ($R \sim 7500$) spectra covering the Ca~{\sc ii} triplet regime (8410--8795\,\AA) of 457\,588 unique stars randomly selected from the southern Hemisphere, in nearly ten years (2003--2013) using the multi-object spectrograph 6dF on the 1.2\,m UK Schmidt Telescope of the Australian Astronomical Observatory (AAO).
The determinations of RVs and atmospheric parameters ($T_{\rm eff}$, log\,$g$, [Fe/H]) for RAVE stars have been detailedly described by Siebert et al. (2011) and Kordopatis et al. (2011, 2013), respectively.

To determine the RVZP of RAVE measurements, we cross match the APOGEE RV standard stars with the RAVE DR5 catalogue (Kunder et al. 2017) and find 352 common stars of RAVE spectral SNR greater than 10.
The stars yield a mean difference $\Delta {\rm RV} = 0.17$\,km\,s$^{-1}$ and a standard deviation s.d.\,$= 1.27$\,km\,s$^{-1}$.
{The standard deviation reveals the RAVE RV measurement errors and is indeed in good agreement with the predicted uncertainty of the RAVE RV in a mean SNR of 48 (Steinmetz et al. 2006).}  
These values are also consistent with the results reported in the previous studies (e.g. Kunder et al. 2017).
The RV differences between RAVE stars  and APOGEE RV standard stars ($\Delta {\rm RV}$) as a function of $T_{\rm eff}$, log\,$g$, [Fe/H] and SNR are also shown in the top panel of Fig.\,8 and no significant trends are detected for those parameters.

\textbf{(ii) The LAMOST Galactic Spectroscopic Survey:} The survey has hitherto collected nearly eight million (currently the largest spectral database) low-resolution ($R \sim 1800$) optical ($3700$-$9000$\,\AA) spectra of SNRs\,$\ge$\,$10$ during its first  five-year Regular Survey (2012--2017) using the LAMOST telescope at Xinglong Observatory.
LAMOST is a 4\,m, quasi-meridian reflecting Schmidt telescope equipped with 4000 fibers distributed in a field of view of 5$^{\circ}$ in diameter (Cui et al. 2012).
Details about target selections and scientific motivations of the survey can be found in Zhao et al. (2012), Deng et al. (2012) and Liu et al. (2014).
Currently, two stellar parameter pipelines, the LAMOST Stellar Parameter at Peking University (LSP3; Xiang et al. 2015, 2017) and the LAMOST Stellar Parameter Pipeline (LASP; Luo et al. 2015), have been developed to derive RVs and basic atmospheric parameters from the collected spectra.

To examine the RVZPs of LAMOST measurements yielded by LSP3 and LASP, we cross match the APOGEE RV standard stars with LAMOST DR3\footnote{Here, the DR3 yielded by LASP contains the data collected between 2011--2015 and can be found at the website: http://dr3.lamost.org/catalogue. The DR3 yielded by LSP3 contains data collected between 2011--2017 and will be publicly available soon (Huang et al. in preparation).} delivered by LSP3 and LASP, respectively.
For the catalog yielded by LSP3, a total of 4849 common stars are found.
The mean RV difference is $\Delta {\rm RV} = -2.60$\,km\,s$^{-1}$, with a standard deviation s.d.\,$=$\,$3.86$\,km\,s$^{-1}$.
The RV differences as a function of $T_{\rm eff}$, log\,$g$, [Fe/H] and SNR are again shown in the middle panel of Fig.\,8. and show almost a constant value for those parameters.
For the catalog yielded by LASP, a total of 3290 common stars are found.
The mean RV difference is $\Delta {\rm RV} = -3.92$\,km\,s$^{-1}$, with a standard deviation s.d.\,$=$\,$3.63$\,km\,s$^{-1}$.
{The relatively large constant offsets between the RVs from the LAMOST pipelines (both LSP3 and LASP) and the RVs from the APOGEE reference standard stars are possibly due to the wavelength calibration of LAMOST spectra.
The current RVs from LAMOST are derived with blue spectra while few sky emission lines are available to recalibrate the wavelength calibration in the blue range.
The $\sim 1$\,km\,s$^{-1}$ offset of RVs between LSP3 and LASP is mainly due to the different masks and wavelength ranges used in the RV determinations.}
As shown in the bottom panel of Fig.\,8, the RV differences $\Delta {\rm RV}$ show no obvious changes as a function of $T_{\rm eff}$, log\,$g$, [Fe/H] and SNR.

{We note that, given the relatively large random errors, the effective temperature/color dependence systematics are not significant for both RAVE and LAMOST RVs , examined by the APOGEE RV reference stars.}

{
\textbf{(iii) The Gaia-RVS Survey:} 
In Gaia DR2 (Gaia Collaboration et al. 2018a), median radial velocities for 7,224,631 stars with $3550 \leq T_{\rm eff} \leq 6900$\,K  and $G_{\rm RVS} \leq 12$\,mag (i.e. $V \leq 13$\,mag) are delivered (Katz et al. 2018; Sartoretti et al. 2018), using the spectra collected by the {\it Radial velocity Spectrometer} (RVS) instrument on-board Gaia.
The RVS is a medium resolving power ($R \sim$\,11,500) integral field spectrograph, covering the   Ca~{\sc ii} triplet regime (845--872\,mm). 

Again, to examine the RVZP of Gaia-RVS measurements, we cross match the APOGEE RV standard stars with Gaia DR2 (Gaia Collaboration et al. 2018a) and find 10,674 common stars.
The mean difference found by these stars is $\Delta$\,RV\,$= 0.36$\,km\,s$^{-1}$, with a standard deviation s.d.\,$= 0.59$\,km\,s$^{-1}$.
The results are in good agreement with the results reported in Katz et al. (2018).
More detailedly, we show the RV differences as a function of color ($G_{\rm BP} - G_{\rm RP}$), $G$ magnitude and number of transits ($N_{\rm obs}$) in Fig.\,9.
The differences are almost a constant value for color and number of transits, but show a clear positive trend of $G$\,magnitude.
In the bright range ($G \leq 11$\,mag), the mean difference is close to zero and then increase to $\sim$\,500-600\,m\,s$^{-1}$ at the faint end ($G \sim 14$\,mag).
This trend is also found by Katz et al. (2018).
The overall offset reported in Table\,3 is mainly due to the large number of faint stars in the common stars. 

In addition, given the high-precision of the RV measurements of our APOGEE RV reference stars, we can quantitatively study the precision of Gaia-RVS RV measurements.
Doing so, we divide the common stars into different metallicity bins, i.e. [Fe/H]\,$\ge -0.1$, $-0.4 \leq$\,[Fe/H]\,$< -0.1$ and [Fe/H]\,$< - 0.4$.
We do not have bins in surface gravity since most of our RV standard stars are giant stars.
For each metallicity bin, we calculate the standard deviation using the common stars as a function of $G$\,magnitude bins, for different effective temperature ranges.
The results are shown in Fig.\,10.
Generally, the precision of Gaia-RVS RVs decreases with magnitude and effective temperature, increases with metallicity, which are all expected.
The typical values of the precision are few hundred m\,s$^{-1}$ at the bright range ($G \sim$\,9-10\,mag) and few km\,s$^{-1}$ at the faint end ($G \sim$\,13-14\,mag).}

\section{Summary}
Using data derived from about one million high-resolution ($R \sim$\,$22$\,$500$) near-infrared ($H$ band; 1.51--1.70\,$\mu$m) spectra of around thirty thousand unique stars released in the SDSS DR14, we have built a catalogue of 18\,080 RV standard stars.
The RVZP of APOGEE instrument is well calibrated to a reference catalog of RV standard stars selected from existing databases.
These APOGEE RV standard stars are observed at least three times and have a stability ($3\sigma_{\rm RV}$) better than 600\,m\,s$^{-1}$ over a time baseline longer than 200 days.
The spatial coverage of these stars is currently limited to the northern sky.
Once more measurements based on the observations of APOGEE-2 at LCO are released, more southern RV standard stars should become available in the future.
Due to the target selections of APOGEE, most of the APOGEE RV standard stars are red giant stars ($J - K_{\rm s} \ge 0.5$).
As add-on sources, about ten per cent blue main-sequence standard stars are also included.
The $H$ band magnitude range is $7$--$12.5$\,mag and the faint limit is much fainter than the previous RV standard stars.

As an application, we have determined the RVZPs of three large-scale stellar spectroscopic surveys: the RAVE, the LAMOST Spectroscopic Surveys {and the Gaia-RVS survey} using this new catalogue of APOGEE RV standard stars.
By comparing the RVs of APOGEE RV standard stars with these yielded by the RAVE and LAMOST pipelines, a negligible offset ($0.17$\,km\,s$^{-1}$)  is found for RAVE RV measurements. 
For LAMOST RVs yielded by LSP3 and LASP pipelines, the offsets are $-2.60$\,km\,s$^{-1}$ and $-3.92$\,km\,s$^{-1}$, respectively.
{The offsets  possibly come from the wavelength calibration of LAMOST spectra.} 
No obvious variations of velocity measurement differences as a function of $T_{\rm eff}$, log\,$g$, [Fe/H] and SNR are found in all cases.
{For Gaia-RVS RVs, the global offset is around 0.36\,km\,s$^{-1}$, with a standard deviation of $0.59$\,km\,s$^{-1}$.
No significant trends of the offsets as a function of color ($G_{\rm BP} - G_{\rm RP}$) and number of transits ($N_{\rm obs}$), but a clear positive trend is found for $G$\,magnitude.
The offset is close to zero at the bright range ($G \sim$\,9-10\,mag)  and reaches $\sim$\,500-600\,m\,s$^{-1}$ at the faint end ($G \sim 14$\,mag).
In addition, we quantitatively study the precision of Gaia-RVS RV measurements as a function of metallicity, effective temperature and magnitude.
The results are consistent with the predicted performances of Gaia-RVS.}

{Currently, our catalogue of APOGEE RV standard stars still has two drawbacks: 1) most of the stars are giants and 2) the spatial coverage is largely in the northern sky.
 In future, with more observations collected by APOGEE-2 and other high resolution spectroscopic surveys (e.g. HERMES-GALAH), we can improve the catalogue to has more main-sequence stars and an all-sky coverage.}

To conclude, the catalogue of APOGEE RV standard stars constructed in the current work should be very useful to determine RVZPs of currently ongoing or future large-scale stellar spectroscopic surveys.

 \section*{Acknowledgements} 
 This work is supported by the National Key Basic Research Program of China 2014CB845700, the China Postdoctoral Science Foundation 2016M600849 and the National Natural Science Foundation of China U1531244 and 11473001.  
The LAMOST FELLOWSHIP is supported by Special fund for Advanced Users, budgeted and administrated by Center for Astronomical Mega-Science, Chinese Academy of Sciences (CAMS). 
 
The Guoshoujing Telescope (the Large Sky Area Multi-Object Fiber Spectroscopic Telescope, LAMOST) is a National Major Scientific Project built by the Chinese Academy of Sciences. Funding for the project has been provided by the National Development and Reform Commission. LAMOST is operated and managed by the National Astronomical Observatories, Chinese Academy of Sciences.

This work has made use of data from the European Space Agency (ESA) mission Gaia (https://www.cosmos.esa.int/gaia), processed by the Gaia Data Processing and Analysis Consortium (DPAC, https://www.cosmos.esa.int/web/gaia/dpac/consortium).

\end{document}